# Securing the IEEE 802.16 OFDM WiMAX PHYSICAL AND MAC Layer Using STBC Coding and Encryption

[1]Samuel Erskine
[1]serskin@bridgeport.edu

[2]Dr. Ziengpieng WU
[2]zhengpi@bridgeport.edu

Department of Computer Science and Engineering
University of Bridgeport, CT, USA

**Abstract:** -This work proposes model design in securing the IEEE 802.16 WiMAX Physical and MAC layer, using Orthogonal Frequency Division Multiplexing (OFDM) and STBC model. Typically, it addresses the physical and MAC layer security concerns, using a Space Time Block Coding (STBC), link encryption, and Message Authentication Code (MAC) technique. The model conforms to Multiple Input Single Output (MISO) fading channels which model two or more transmitters and a receiver in multiuser environment.

The two fading link parameters are assumed to be same. Channel estimate for each link, in combination to the received signal is based on Reed Solomon Convolution Coding (RS-CC) algorithm, which occurs as a result of the Space-Time Diversity Combiner block. In addition the model explore using communication blocks to measure and display bit error rate after encryption algorithm and Message Authentication Code (MAC) have been adapted in Forward Error Correction (FEC) mode. Channel SNR and estimation in rate ID is applied. The final results shows authentication, and the Reed-Solomon decoding of the final information or data received.

1. INTRODUCTION

Further security measures are required for the IEEE 802.16 OFDM WiMAX Physical layers by modern multiplexing and encryption coding techniques. Despite the fact that, researchers have acknowledged that the new WiMAX standard includes a lot of security enhancements, increasing security challenges are presented in its protocol layer design [1]. Secure coding security challenges, such as cryptography integrations, and support are examples. These security issues are normally identified with the WiMAX communication as inherent openness [2]. Moreover, since the WiMAX belongs to the wireless group [3], and because of susceptibility issues, security must be considered as urgent need to the network users and the providers as well.

Furthermore, there are security issues such as rouge base station, and more of eavesdropping attacks, DoS, replay and other modification attacks [4]. Rouge base station, burst communication symbols, are usually identified with the WiMAX channels. Security issues with regard to the WiMAX burst communication symbols, during multiuser communication situation are important for investigation. A novel model, known in the literature as Space-Time Block Coding (STBC), and cryptographic principles such as encryption and Message Authentication Code is desired which strengthen the security issues of such as confidentiality of the IEEE 802.16 WiMAX physical and MAC layers [4].

The rest of the work shall be organized as follows: Section 2 discusses the literature review. In Section 3, we discuss about motivation in WiMAX. Section 4: is the model design of the WiMAX IEEE 802.16 and STBC.

## 2. RELATED WORK

The literature study reviews aspect of communication in IEEE 802.16 WiMAX.CDMA offers strong resistant in preventing against many vulnerable security attacks in the communication of WiMAX channel. However, CDMA is yet identified with eavesdropping security vulnerabilities.[5] used Performance Analysis of a LDPC Coded CDMA System with Physical Layer Security Enhancement" to increase the in-built security of the physical layer transceiver, using secure scrambling process, but achieved minimal bit error rate and high performance in the receiver. Performance assessment, including other secure coding methods were neglected. This work considers it necessary to investigate appropriate measures in secure coding technique that can be implement access to real performance.

[6] used "Physical Layer Built-In Security Analysis and Enhancement of CDMA Systems". This CDMA system is used in military and other commercial applications. CDMA used as 3G has appropriately solved many security concerns such as jamming and eavesdropping. Subsequently, efficient data delivery in the MAC and physical layer has been possible. Although, the implementation of the physical layer in-built security enhancement of the CDMA system application in [6] used encryption based security known as scrambling. A typical concern of how application can secure the IEEE 802.16 WiMAX transmissions is urgently required.    Secure Physical Layer using Dynamic Permutations in Cognitive OFDM Systems technology for security at the physical layer was suggested by [7]. By this, user's data symbols were mapped over the subcarriers by the use of permutation formula for security reasons [7]. The intention was to resolve the physical layer with random and dynamic subcarrier. However, there was multiuser transceiver issue which needs further investigation.

[8] used Wireless Network Token-Based Fast Authentication. Based upon this, WiMAX and Wi-Fi were recognized as important and popular technology to support the wired and wireless network seamless mobility. However, some restriction applies, based on inherent openness. Hence, a security with Medium Access Control (MAC) and the physical layer is evidently clear. Since information must be error free, immediate solution to efficient data delivery in physical and the MAC layers are required.

## 3. MOTIVATION

There is much inherent openness with the IEEE 802.16 WiMAX. By [9], the wireless security openness assessment is based on jamming. Scramble also presents another security challenge. Ancient White Gaussian Noise (AWGN) is usually presented in multipath fading channel. Multipath channel fading propagation of noise in the channel is critical. Overall security solution such as encryption and Message Authentication Code (MAC), is required. The solution to this is realized in the physical and MAC layer protocol design using the STBC. This work essentially investigates the implementation of the security enhancements method of the protocol layer design in STBC. The onus task is that,  channel jamming and scramble at the MAC and physical layer present a greater security challenge. Thus any attacker or masquerader can serve as potential security vulnerability. Moreover, legitimate network users do not have full right to access network resources.

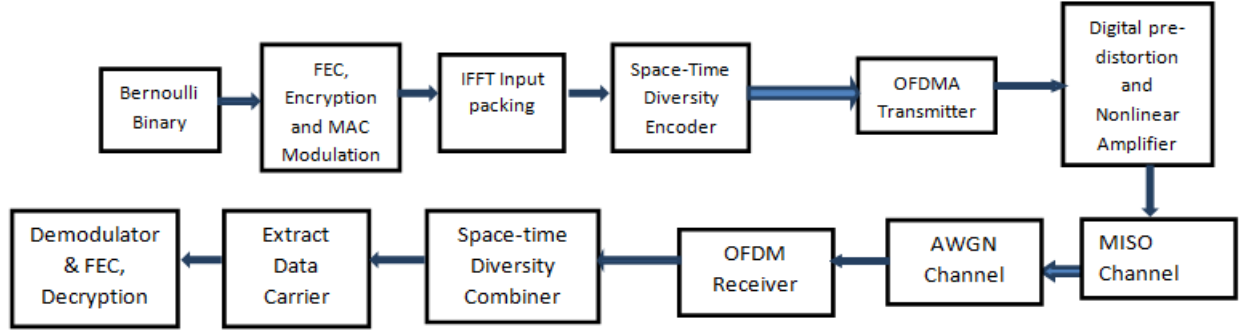

Figure 1: Model Architecture for the IEEE 802.16 OFDM WiMAX Physical Layer

## 4. MODEL DESIGN OF IEEE 802.16 OFDM & SPACE TIME BLOCK CODING

The model design incorporates components in Space-Time Block-Coding (STBC) as:

- Generation of random data bits, which models downlink burst of OFDM integer symbol numbers
- Forward Error Correction (FEC) which uses concatenated codes with rate-compatible convolution code. The rate compatible convolution codes uses encryption and Message Authentication Code (MAC) modeled in the FEC, to secure final received data information
- Use of data interleaving
- Modulation techniques which uses the 16, 32 and 64-QAM constellation
- OFDM transmissions consisting of 192 8 pilots, 256-point FFTs and cyclic prefix length variable sub-carriers
- STBC model transmitter that transmits the only single preamble. Also, existing is the optional pre-distortion with capability to correct nonlinearity in the communication model.
- Multiple-Input Single-Output (MISO) multipath fading channel that has the AWGN received in the STBC model
- Capability and choice of non-fading, flat fading, dispersive multipath fading channel configurations to be non-fading.
- The receiver uses OFDM that include channel estimation with the preamble inserted.
- Based upon this the STBC model diversity combination of the Reed Solomon Convolution Coding (RS-CC) algorithm is important [10].

## 5. STBC LGORITHM USING REED SOLOMON CONVOLUTION CODING (RS-CC)

We now formulate the decoding problem of the final coded information as follows:
Let us represent a codeword $C = C_0 C_1, \ldots C_{n-1}$ as a polynomial codeword which is defined by $c(x)$.
Suppose that $c(x)$ is transmitted, and the polynomial received from this is $r(x)$. The polynomial error can be defined as $e(x) = r(x) - c(x)$
As observed, wherever transmission has occurred, this polynomial will have a non-zero coefficient in its positions.
We now define a problem to find most probable polynomial error $e(x)$, $given\ r(x)$. Let us also try to recall the most probable error pattern that has minimum Hamming weight. Now Define
$S_i = e(\propto^i), i = 1, 2, \ldots n - 1$ (1)
Observing every codeword having $\propto^i, i = 1, 2, \ldots 2t$ serves as the root and also as $r = c + e$ for codeword $c$, we will have

$S_i = r(\alpha^i), i = 1, 2, \ldots 2t$  (2)

Now supposing no error exist, then all $2t$ in above defined syndrome would occur as zero. Meaning, errors will only occur in the transmissions whenever it is realized that one or more of these syndromes shows as non-zero.

Restating the decoding problem:
Given that $S_1, S_2 \ldots S_{2t}$ *then also find* $e(x)$
This indicates decoding problem is more or less like interpolation.

Now, we represent the Hamming weight of error pattern *e to be at most t*. This indicates the error pattern has at most *t non − zero* components. We also recall from the fact that we are guaranteed exactly one 'closest' codeword.

In the digit $e_k$ (where subscript of e starts from 0) is from $j^{th}$ non-zero component, counting from the left, we define $X_j = \alpha^k$

Let's take for instance, $n = 7. It\ is\ expedient\ to\ write\ e(x) as$
$e(x) = e_0 + e_1 x + e_2 x^2 + \cdots + e_6 x^6$. This means the first non-zero coefficient and the third non-zero coefficient are respectively $e_1, and\ e_3$, supposing the first and the third coefficients of non-zero. We deduce that $X_1 = \alpha^1$ and $X_2 = \alpha^3$, here $\alpha$ generates the $n$ order cyclic group which occur in multiplicative group $GF(2^m)$. When $\alpha^j$ are believed to be known, then it will be possible to also find $j$, such that $X_j$ will be known as the error locations. The decoding problem will not completely be solved; the magnitude of the errors will have to be determined. Now the error magnitude $Y_j$ at location $X_j$ will be found as $e_k$, yet we mentioned earlier that $e_k$ is non-zero $j^{th}$ component of the error vector.

By definitions of $X_j, and\ Y_j$, are based on the assumption we made that at most *t errors have* occurred from equation (2):
$S_i = \sum_{j=1}^{t} Y_j X_j^i,\ i = 1. 2, \ldots n - 1.$ (3)
We define the series
$\frac{1}{1-X_j x} = 1 + X_j x + X_j^2 x^2 + \cdots$ (4)
Multiply both side of equation (4) by $Y_j X_j$, also summing both sides from 1 to $t$, using equation (3) we get
$\sum_{j=1}^{t} \frac{Y_j X_j}{1-X_j x} = S_1 + S_2 x + \cdots . S_{2t} x^{2t} + \cdots$  (5)
The terms of left hand side of equation (5) is combined to form a single fraction as
$\frac{\sum_{j=1}^{t} Y_j X_j \prod_{k=1, k \neq j}^{t}(1-X_k x)}{\prod_{j=1}^{t}(1-X_j x)}$ (6)
The numerator and denominator of equation (6) are replaced by $w(x)$ and $\sigma(x)$ respectively. We can then write equation (5) as
$\frac{w(x)}{\sigma(x)} = S_1 + S_2 x + S_3 x^2 + \cdots S_{2t} x^{2t-1} + \ldots$ (7) Hence, we deduce the following properties of equation (7):

1. The left hand side denominator $\sigma(x)$ has degree one greater than that of numerator.
2. The error inverse locations happen to be the polynomial root in the denominator. Hence the polynomial is termed error locator polynomial.
3. The left hand side numerator and the denominator polynomials are relatively prime. In checking to verify this assertion, it can be observed that none of the denominator roots happens to be the numerator roots [11].

## 6. ENCRYPTION AND DECRYPTION IN WiMAX USING CIPHER BLOCK CHAINING (CBC)

The IEEE 806.16 WiMAX has security deficiencies. WiMAX security has suffered the same faith, due to much inherent openness. Cipher Block Chaining (CBC) given in figure 2 [13] model has been beneficial to secure most of this openness. Block protocol layer communication in the model design as in Figure1 has been suggested for improvement. This model takes advantage of FEC transmissions that include encryption and Message Authentication Code (MAC) of same plaintext block. The encryption is repeated, which also produces varied cipher text blocks. A cipher block chaining (CBC) implementation model is shown in figure 2 below as requirement [12]

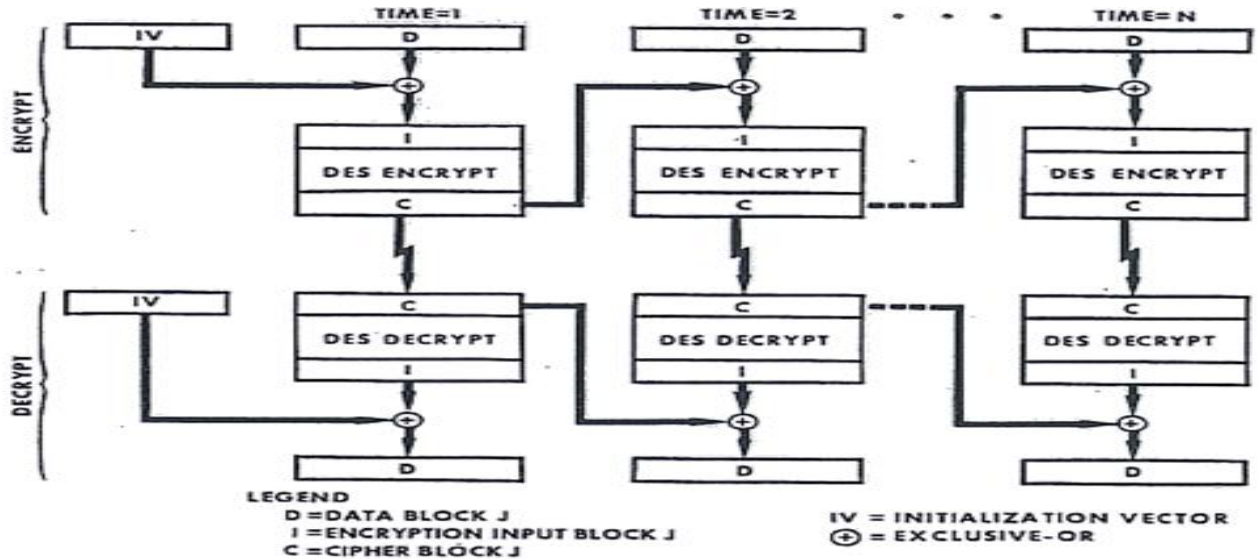

Figure 2. Cipher Block Chaining Implementation model Encryption and Decryption

The scheme is designed in such a way that encryption algorithm input is the current plaintext block XOR function, including the preceding ciphertext block which uses same key for each block. Eventually, sequence of plaintext chaining has been possible together. Each plaintext block input of the encryption function does not have any relationship to the plaintext block. This has made it possible for repeating *b* bits patterns which are not being exposed.

The decryption algorithms consist of each cipher block being passed through the decryption process. In order to produce the plaintext back, the preceding ciphertext block is XORed with the result obtained. Hence, in order to observe this working the following formula can be deduced.

$$C_j = E(K[C_{j-1} \oplus P_j])$$

Then, $D(K, C_j) = D(K, E[(C_{j-1} \oplus P_j])$

$$D(K, C_j) = C_{j-1} \oplus P_j$$

$$C_{j-1} \oplus D(K, C_j) = C_{j-1} \oplus C_{j-1} \oplus P_j = P_j$$

In order to generate first block of ciphertext, the first plaintext block is XoRed with an initialization vector (IV). The first plaintext block can be recovered on decryption by XoRing IV with the decryption algorithm output. The size of the Initialization Vector (IV) block represents same size as the cipher block.

In this model the transmitter and the receiver are conversant with the IV; any third party which tries to identify with the IV must be identified as security vulnerability, and the system should prevent any attack with that. In order for maximum security protection to be realized, the IV should be protected by using authentication and confidentiality to prevent any unneeded changes. Electronic Code Book (ECB), whose description is beyond this work scope, is an encryption technique that can help send the IV; this is to help this encryption scheme which include authentication to be realized. Several reasons are necessary in which one can think of securing the IV. An urgent reason could be that if the receiver is tricked by a jammer or scrambler, or any security vulnerability such as noise, using diverse IV as cryptanalytic attack or brute force, then the bits selected, could possibly be inverted in the initial plaintext block. However the system is able to control. Let us observe this taken place: $C_1 = E(K, [IV \oplus P_1])$

$$P_1 = IV \oplus D(K, C_1)$$

Now, the $ith$ bit in the $b - bit$ in quantity X is denoted by $X[i]$. Therefore,

$$P_1[i] = IV[i] \oplus D(K, C_1)[i]$$

Hence by using XOR properties, it is important to note that

$$P_1[i]' = IV[i]' \oplus D(K, C_1)[i]$$

Based on the interpretation of this, the notation of the prime denotes bit complement. Also, an opponent who can truly predict bit change in the IV would encounter the correspondent received value bit $P_1$ possibly as changed [12].

### 6.1. Data Encryption Standard (DES) and Message Authentication Code (MAC)

In the IEEE 802.16 WiMAX, Data Encryption Standards (DES), and Data Authentication Algorithm (DAA) represent one of the possible means by which secure data protection practices can be adapted. This is based on exceptionally use of Message Authentication Code (MAC) for some time now [13]. Though, some security weaknesses concerning the use of the algorithm have been detected [13], a newer and stronger algorithm ought to be proposed. This newer proposed model algorithm is necessary in the IEEE 802.16 OFDM WiMAX physical layer, which is related to the proposed Cipher Block Chaining (CBC) (Figure 3). Based upon that, describing a model operation in DES becomes a reality, which also uses zero initialization vector (IV). Information or data may appear in many forms such as message, program or a particular file. This ought to be authenticated and must be organized into 64-bit contiguous blocks such as: $D_1, D_2, D_3, \ldots D_n$. We assume padded bit are needed in the model design.

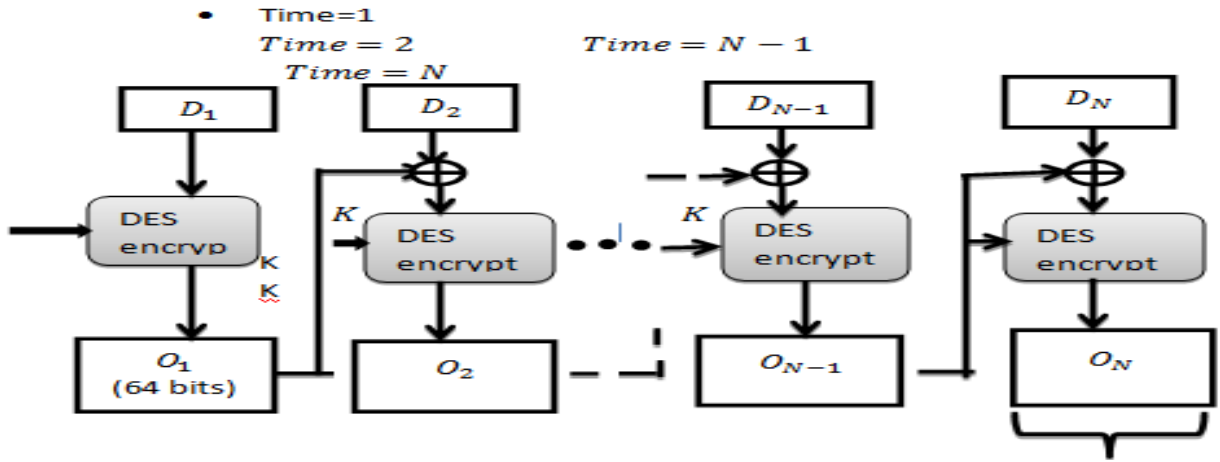

Figure 3: Proposed WiMAX Frame Authentication

This forms a complete 64-bit block, which is linked to Forward Error Correction (FEC) – see model design shown in Figure 1. Over here, DES encryption algorithm is enforced that has $E$, a secret key $K$, and data authentication code (DAC) derived in the following as:

$$O_1 = E(K, D_1)$$

$$O_2 = E(K, [D_2 \oplus O_1])$$

$$O_3 = E(K, [D_3 \oplus O_2])$$

- 
- 
- 

$$O_N = E(K, [D_N \oplus O_{N-1}])$$

This indicates that the entire block $O_N$, or the blocks leftmost M bits with 16≤ M ≤ 64 implement DAC [12].

### 6.2. Model Simulation of the IEEE 802.16 WiMAX

The model parameters for simulation comprises such as shown in tables 1 and 2.

Signal constellations with signal points coordinates with constellation points as: 16QAM, 32QAM, and 64QAM are modulation techniques of the OFDM which models burst communication symbols of the WiMAX. Security requirements which include encryption and authentication to provide confidentiality. In table 2 also shows the model parameters and the corresponding values. Initial seed requirement which concern random positive integer of the OFDM were considered. The mode and its value being the BER. The $E_o/N_o$ (dB) and its variables in range were used in the BER tool analysis. The input signal power (Watts) considered is 1W. The symbol per period and their values one (1) per each symbol rate were used.

**Table 1: Parameter Settings for QAM Simulation**

| Parameter | Value |
|---|---|
| Signal Constellation | Signal points coordinates in constellations(16,32, 64) |
| Samples per symbols used | 2 |
| Security | Confidentiality, Encryption, Authentication |

**Table 2: Parameter Setting For AWGN**

| Parameter | Value |
|---|---|
| Initial seed | Random positive integer |
| Mode | BER |
| $E_o/N_o$ (dB) | Variable |
| Input Signal power (Watts) | 1 |
| Symbol per period | 1/symbol rate |

## 7. RESULT DISCUSSION

The following can be deduced from the simulation graph using Matlab/Simulink:

Figure 4 uses theoretical simulation analysis of BER QAM at 16, 32, and 64 constellation points of that model downlink burst of data in the OFDM. It computes the bit error rate (BER) of the signals received. The final received signals are encrypted and include Message Authentication Code (MAC). The graph indicates as $E_o/N_o$ is increasing to the right, the BER diminishes, and less error occurs in the receiver.

Figure 5 uses semianalytic simulation. Based upon this, BER is assessed at the QAM at the 16, 32, and 64 constellation points modeling OFDM burst downlink data of received signals. Error probabilities of the received signal are determined. The ratio $E_o/N_o$ (dB) representing overall received signal encrypted and authenticated is increasing to the right, the BER decreases, indicating less probable error in all three simulations.

Figure 6 represent four Monte Carol simulations graphs. Based upon this, BER represents bit energy per symbols, each of which has noise power spectral density $E_o/N_o$ levels. It uses the function viterbisim which uses coding of the OFDM burst data to computes the overall system error rates. Based upon each simulations graph, the points in each graphs are closed to each other in the four simulations. The results is such that decoded received signal is encrypted and authenticated, confidentiality is assured. The ratio $E_o/N_o$ as increasing to the right, each of the four simulation points indicates less BER.

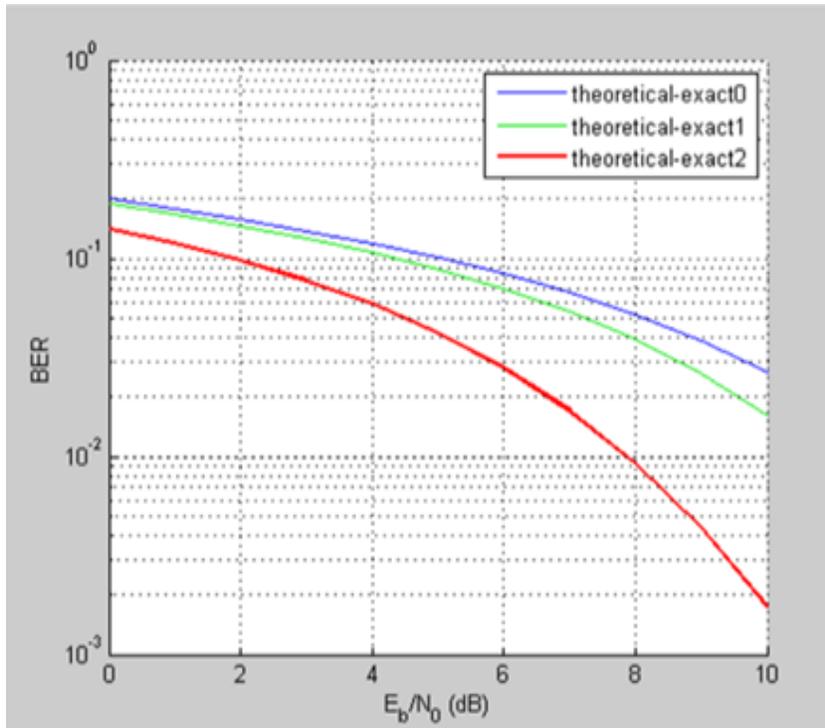

Figure 4: Plot of Theoretical BER over QAM

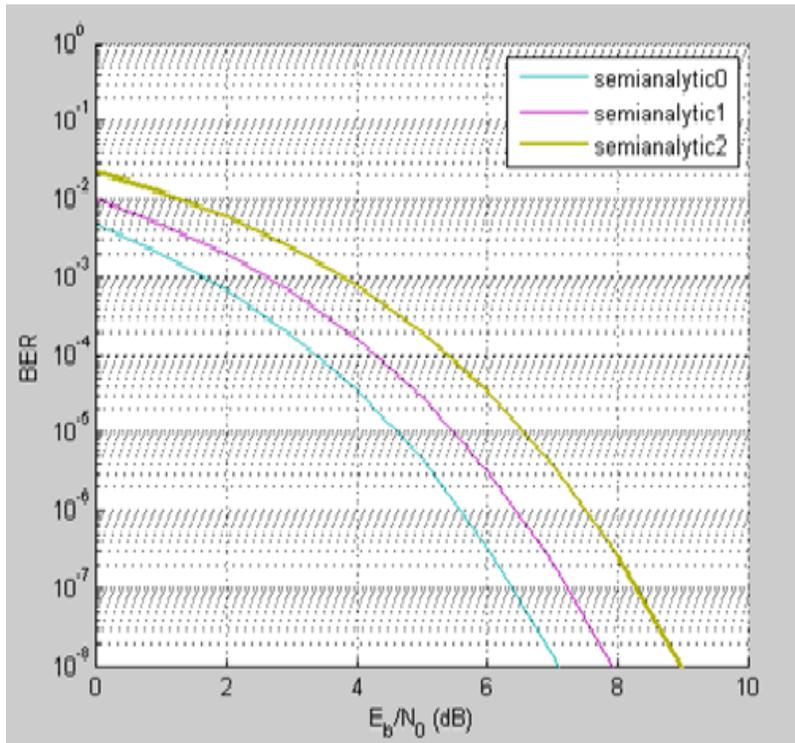

Figure 5: Plot of Semi analytic BER over QAM

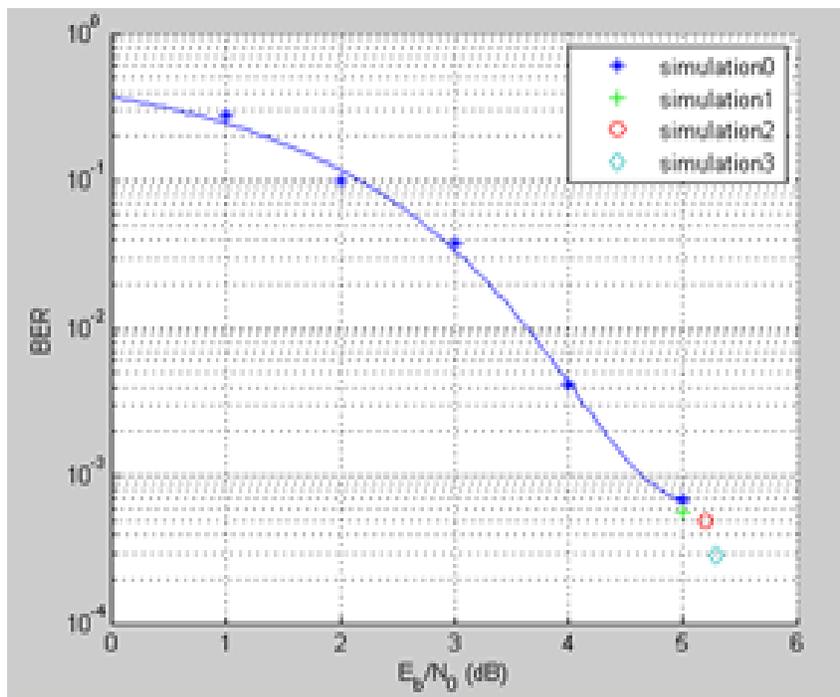

Figure 6: Plot of BER over Bit Energy per Symbol to Noise Spectral Density

## 8. CONCLUSION

It has often, been erroneously imagined that, only security devices are capable of ensuring secure end-to-end 3G WiMAX. All aspect of security consideration must also be considered. Let take for instance secure coding, and cryptology. Confidentiality is an issue in many wireless mediums. These have often been neglected in standard technology such as WiMAX [13] (recommendations from WiMAX forums). Confidentiality, encryption and Message Authentication are often assumed. Confidentiality need to be detailed in algorithms. Also important security technology is coding. Space Time Block Coding (STBC) algorithms must be integrated into all wireless devices.

We have detailed in our work an algorithm using Reed Solomon Convolution Coding using STBC. Simulation result has proved the effectiveness of the algorithm, providing message confidentiality, authentication and encryption.Consequently, we have secured the WiMAX has based upon the Space Time Block Coding (STBC) and encryption techniques, including the use of message authentication code (MAC) for the received signals have been possible. The numbers of the OFDM symbols are set constant including burst data generated. This cause all profiles frame duration in the simulation to remain same. The simulation has effectively modeled burst downlink data. In future work, IP transport mechanism will be considered in al downlink modeling of WiMAX burst data at the base station.

**References**

[1] P. Rengaraju, C. Lung, and A. Srivasana (2010). "Design of Distributed Security Architecture for Multihop WiMAX Networks". Private Security and Trust (PST) 2010 IEEE International Conference. Pages 54-61
[2] P. Regaraju, C. Lung, S. Srivasana and Y. Qu (2009). "Analysis of Mobile WiMAX Security". Science and Technology for humanity (TIC-STH), 2009 IEEE Toronto International Conference. Pages 439
[3] P. Rengaraju, C. Lung, and A. Srivasana (2011). "Measuring and Analyzing WiMAX Security and QoS in Testbed Experiments". Communications (ICC), 2011 IEEE International Conferences. Pages 1-5

[4] IEEE P802.16m/D4, "DRAFT Amendment to IEEE Standard for Local and metropolitan area networks—Part 16: Air Interface for Broadband Wireless Access Systems" (2012). 10.1109/IEEESTD.2012.6272299 Publication Year: 2012, Page 1

[5] T. Shojaeez and, Dr. P. Azmi, and A. Yadegari, (2010). "Performance Analysis of a LDPC Coded CDMA System with Physical Layer Security Enhancements". IEEE 2010 Publications Conference. Pages 1 and 4.

[6] T. Li, J. Ren, Q. Ling and A. Jain. "Physical Layer Built-In Security Analysis and Enhancement of CDMA Systems". Military Communication Conference, 2005. MILCOM 2005 IEEE. Pages 1 and 7.

[7] F. Meucci, S. Wardana, and N. Rashmi, (2009). "Secure Physical Layer Using Dynamic Permutations in Cognitive OFDMA Systems". IEEE 2009 Conference. Pages 1 and 5.

[8] G. Kbar, (2010). "Wireless Network Token-Based Fast Authentication". IEEE 2010 17th International Conference on Telecommunications. Pages 227 and 232.

[9] M. Bloch (2008) "PHYSICAL-LAYER SECURITY". A Dissertation. Pages 6

[10] Mathworks, Matlab and Simulink Software package

[11] P.Shankar (2007). "Decoding Reed Solomon Codes Using Euclid's Algorithm" http://www.researchgate.net/publication.

[12] W. Stallings (2006)."Cryptology and Network Security".Block Cipher Mode of Operation. Pages 183-185, 383.

[13] WiMAX Security forf Real World Network Service Deployment. http://www.motorola.com/web/Business/Products/Wireless.

**Authors Biography**

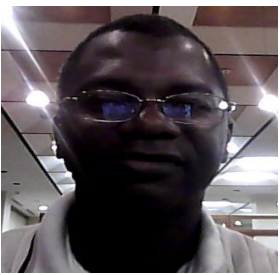

Samuel Kofi Erskine is a PhD student and Graduate research assistant at Department of Computer science and Engineering University of Bridgeport CT USA. He obtained his Masters of Science in Telecommunications at George Mason University, Virginia USA. Samuel has authored and published two research papers in Computer Science & Engineering and business Technology management in international journals of repute.

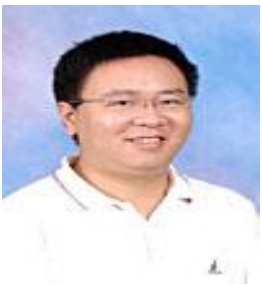

Dr. Zhengpiw Wu is assistant Professor of Computer science and Engineering at Department of Computer Science and Engineering at University of Bridgeport. He obtained his PhD at University of Virginia USA, and Bachelor of Engineering (BEng) in Computer Science at Zhejiang University in China. Dr. Wu's Research Interests lies in: Distributed System, Information Security, Web Service,E-Commerce, Healthcare Informatics, Service-oriented Computing and Grid Computing, . Networking, Knowledge Engineering, Fuzzy Logic. He has authored and published a lot of papers in computer Science and Engineering in journals of repute.